\documentclass[table, xcdraw, fleqn,10pt]{wlscirep}
\usepackage{xcolor}
\usepackage[utf8]{inputenc}
\usepackage[T1]{fontenc}
\usepackage{graphicx}
\usepackage{blindtext} 
\usepackage{algorithm}
\usepackage{algpseudocode}
\usepackage{diagbox}
 \usepackage{multirow, balance}

\usepackage{stfloats, microtype} 

\usepackage[english]{babel} 

\usepackage{hyperref} 

\usepackage[caption=false]{subfig}
\usepackage{lipsum} 
\usepackage{balance}

\title{Modeling Visual Hallucination: \\A Generative Adversarial Network Framework}

\author[1]{Masoumeh Zareh}
\author[1,2]{Mohammad Hossein Manshaei}
\author[1,*]{Sayed~Jalal~Zahabi}
\author[3]{Marwan~Krunz}
\affil[1]{Department of Electrical and Computer Engineering, Isfahan University of Technology, Isfahan, 84156-83111, Iran}
\affil[2]{Department of Computer Science, Hunter College, City University of New York, New York, NY, USA}
\affil[3]{Department of Electrical and Computer Engineering, University of Arizona, Tucson, AZ, USA}

\affil[*]{Corresponding Author:  zahabi@iut.ac.ir}


\begin{abstract}
Visual hallucination refers to the perception of recognizable things that are not present. These phenomena are commonly linked to a range of neurological/psychiatric disorders.
Despite ongoing research, the mechanisms through which the visual system generates hallucinations from real-world environments are still not well understood.
Abnormal interactions between different regions of the brain responsible for perception are known to contribute to the occurrence of visual hallucinations.
In this study, we propose and extend a generative neural network-based framework to address challenges within the visual system, aiming to create goal-driven models inspired by neurobiological mechanisms of visual hallucinations.
We focus on the adversarial interactions between the visual system and the frontal lobe regions, proposing the Hallu-GAN model to suggest how these interactions can give rise to visual hallucinations. The architecture of the Hallu-GAN model is based on generative adversarial networks.
Our simulation results indicate that disturbances in the ventral stream can lead to visual hallucinations.
To further analyze the impact of other brain regions on the visual system, we extend the Hallu-GAN model by adding EEG data from individuals. 
This extended model, referred to as Hallu-GAN$^+$, enables the examination of both hallucinating and non-hallucinating states. 
By training the Hallu-GAN$^+$ model with EEG data from an individual with Charles Bonnet syndrome, we demonstrated its utility in analyzing the behavior of those experiencing hallucinations.
Our simulation results confirmed the capability of the proposed model in resembling the visual system in both healthy and hallucinating states.

\end{abstract}
\begin{document}

\flushbottom
\maketitle
%
%
\thispagestyle{empty}


\section{Introduction}

\label{Introduction}
Understanding the brain’s functionality has long been a challenge, drawing significant attention in neuroscience over the past two decades. The brain is an essential body organ for information processing and memory.
In the brain, neurotransmitters serve as a means to connect the different areas, allowing them to interact with each other for information processing~\cite{berg2013neurotransmitter, purves2004neuroscience}.  Certain brain damage can result from neurological illnesses or aging, which can disrupt neurotransmitters and connections between different brain areas. One of the known symptoms of many brain diseases is hallucination, formally defined as the unpredictable experience of perceptions without corresponding sources from the external world~\cite{teeple2009visual}. 
There are five types of hallucinations—auditory, visual, tactile, olfactory, and taste—that can occur in various diseases, including schizophrenia, Parkinson's disease, Alzheimer's disease, migraines, brain tumors, Charles Bonnet syndrome, and epilepsy~\cite{teeple2009visual,silbersweig1995functional, moustafa2016cognitive, ramirez2007cerebral, weil2016visual, guest2020simulating}.


Scientists try to understand the brain by the use of abstract models, especially in the field of computational neuroscience~\cite{o2000computational}. 
So far, researchers have identified several aspects of the brain's structure and functionality with modeling, but the complexity of the brain still presents a significant challenge. 
Models help to dissect and study the abundant processes occurring in the brain, allowing for a more precise analysis of the connections between different brain regions. In the past two decades, Artificial Intelligence (AI) techniques have been applied in uncovering the mysteries of the brain~\cite{tenti2008mathematical,colombo2018discovering, dichio2023statistical, chen2019roles}.
In particular,  numerous studies focused on modeling hallucination and predicting its impact~\cite{muller2014visual,thomas2023changes, hare2021hallucinations, grana2019dynamic, klein2020perceptual, reggia1996computational}.  For example, in~\cite{ grana2019dynamic}, Dynamic Causal Modeling (DCM)  was considered for estimating the effective connection between brain regions on resting-state functional magnetic resonance imaging (rs-fMRI) to analyze auditory hallucination in Schizophrenia patients. Similar to other scientific disciplines, AI has made contributions to enhance our understanding of this phenomenon in the brain. 

Methodologies for modeling hallucincation are divided into four main classes. 
The first class focuses on inferring brain function through mathematical models~\cite{dandi2020generalized,  friston2003learning, gershman2017complex, zareh2019neurons, mohamed2023dynamic}. One of the key common concepts relevant to our study here is inference, which is to ascertain the probability of a potential cause given an observation~\cite{friston2003learning}. 
The second trend involves using Reinforcement Learning (RL) to understand the connections between different brain areas and neuron networks~\cite{girdler2022neural, eckstein2021reinforcement}. 
The third class utilizes deep learning (DL) techniques to model brain processes and detect diseases by analyzing brain data~\cite{yamins2016using, dobs2022brain}. 
The fourth class is a more recent research direction. It considers exploiting generative models, including generative adversarial networks (GANs)  and Diffusion models, in neuroscience. 
Recently, utilizing the idea of GAN~\cite{goodfellow2014generative},  an adversarial framework was proposed for probabilistic computation in the brain~\cite{gershman2019generative, cushing2023generative}. These frameworks demonstrate how communication between the discriminator and the generator of GAN may explain the delusions observed in some mental diseases and mental illnesses such as post-traumatic stress disorder (PTSD)~\cite{gershman2019generative, cushing2023generative}.

In this paper, we look into the evidence within the neurobiology and neuropsychiatry of the human brain aiming at developing a framework for approximate inference in the hallucinating brain. Specifically, we examine how the perception and encoding of visual inputs contribute to the occurrence of hallucinations in certain disorders. This modeling facilitates the comprehension of the operational mechanisms of the visual system and assists in the identification of circumstances that could potentially elicit hallucinations. To illustrate the significance, imagine a psychiatrist trying to prevent a patient from entering a hallucinatory state. How can they effectively recognize such situations?


Generative models are typically employed for generation tasks, such as generating EEG/fMRI data, rather than tackling extensive brain modeling problems. We propose a methodology for employing GANs as an adversarial framework for modeling the hallucination observed in some neurologiacl/mental health conditions, such as Parkinson's disease and Schizophrenia. 
We use an adversarial model similar to~\cite{gershman2019generative, cushing2023generative} for modeling the visual system.  We review the biological aspects of the hallucination process and illustrate how it could occur within the visual system.  
We further investigate the proposed model through simulations, showing the differences in connections between layers during two distinct states (healthy and hallucinatory states) and comparing them with relevant brain states. 
In the next step, we explain the input of our model to demonstrate the influence of memory in this process. Through the use of an actual clinical sample, we illustrate how this feature can impact hallucination.

Computational psychiatrists/neuroscientists often prefer to follow approximate inference by exploring the biological implementation of Monte Carlo and variational methods~\cite{gershman2017complex}.
Inspired by~\cite{gershman2019generative}, our approach relies on an adversarial inference setup, which provides several significant advantages over standard Monte Carlo and variational approaches. 
First, it can be applied to more complex models.
Second, it is more efficient than the standard Monte Carlo algorithms and can use more flexible approximate posteriors compared to standard variational algorithms~\cite{gershman2019generative}.
Third, GAN-based adversarial learning techniques directly learn a generative model to construct high-quality data, making them generally more realistic than variational approaches~\cite{dandi2020generalized}.



The rest of this paper is organized as follows: Section~\ref{sec_Related} presents related work. Section~\ref{sec_pre} provides a preliminary overview of the GAN concept and highlights the relevant evidence within the mechanism of visual hallucinations. Then, we describe in detail the proposed model for visual hallucinations in Section~\ref{HM}. In Section~\ref{Sim_model}, we present the experimental results and an analysis. Finally, we discuss the result of models in Section~\ref{sec_Discussion}, followed by the conclusion presented in Section~\ref{sec_Conclusion}.

\section{
Related Work
}
\label{sec_Related}
Since the mid-twentieth century, there has been significant progress in the principled design and discovery of biologically and physically informed models of neuronal dynamics~\cite{zareh2019neurons,dandi2020generalized,  friston2003learning, gershman2017complex, mohamed2023dynamic}.
Recent developments provide intriguing insights into the use of AI techniques within computational neuroscience~\cite{collerton2023understanding, ramezanian2022generative}.

Following the first class of modeling hallucination mentioned in Section~\ref{Introduction}, in~\cite{gershman2017complex} the neural mechanisms were studied via probabilistic inference methods. The brain's structural and functional systems are seen to possess features of complex networks~\cite{bullmore2009complex}. Additionally, visual hallucinations are largely understood through generative approaches, especially Friston's Active Inference framework~\cite{collerton2024episodic, friston2003learning}. Inverted encoding models were used to infer encoding mechanisms influenced by cognitive states across a wide range of stimuli in vision neuroscience~\cite{hays2024leveraging}. These phenomena occur when hallucinatory expectations surpass actual sensory input. This imbalance occurs as the brain aims to minimize informational free energy, which reflects the gap between predicted and actual sensory data in a stable system~\cite{collerton2023understanding, collerton2024episodic, firbank2024functional}. In Section~\ref{HM}, we outline our model, which is grounded in the framework of visual perception as described in~\cite{collerton2023understanding}.
Considering the second trend of hallucination modeling, the field of computational analysis has increasingly incorporated RL to analyze decision-making processes~\cite{fan2023advanced, andres2024brain, yang2024reliance}. Decision-making is adaptive and sensitive to the neural costs associated with different strategies~\cite{yang2024reliance}.
This approach, which aims to replicate the human learning process through trial-and-error experiences, provides a normative framework for thoroughly exploring decision-making~\cite{niv2009reinforcement}. 

Regarding the third trend of hallucination modeling, deep hierarchical neural networks (DHNNs) were used to study computational sensory systems models, especially the sensory and visual cortex~\cite{yamins2016using,    mohsenzadeh2020emergence}. These results suggest that deep convolutional neural networks (DCNNs) are a good approximation of the perceptual representation generated by biological neural networks~\cite{mohsenzadeh2020emergence, zhuang2021unsupervised, xue2024convolutional}.
Generative models allow for the generation of new synthetic data instances that are statistically similar to the training data in the fourth trend of hallucinating modeling. 
Generative methods (such as Autoencoder, GAN, and Diffusion models) are currently used to reconstruct realistic images and analyze brain activity based on fMRI/EEG data~\cite{tirupattur2018thoughtviz, qiao2020biggan, luo2024brain, al2021reconstructing}. The latest generation of generative models can easily solve these generation tasks without challenges.
Furthermore,  some of these frameworks—GAN in particular—are utilized to describe how brain regions interact with one another in certain mental illnesses. 

An encoder-decoder model was introduced in~\cite{adeli2022brain}, designed to mimic the interacting top-down and bottom-up visual pathways comprising the brain's recognition attention system. In this modeling, the ventral pathway can be mapped to encoder processing, and Decoder processing maps onto the dorsal pathway~\cite{adeli2022brain}. 
In ~\cite{cushing2023generative}, it was suggested that the subjective experience of visual imagination depends on GAN-like mechanisms, and the authors investigated whether this approach can help us better understand the intrusive imagery experiences of people with mental conditions such as PTSD and acute stress disorder. From another viewpoint, it is believed that when earlier experiences are systemically replayed, offline states like sleep increase the ability of humans and other animals to derive general concepts from their sensory experiences~\cite{deperrois2023learning}. 
In~\cite{deperrois2022learning}, a model called the perturbed and adversarial dreaming mode (PAD) based on the functional cortical architecture was developed. The model uses GANs to implement adversarial learning in cortical circuits and their plasticity mechanisms. 
It suggests that perturbed dreaming during non-rapid eye movement (NREM) sleep and adversarial dreaming during rapid eye movement (REM) sleep can affect learning.  
Each state optimizes a different objective function, but they work together in a complementary manner~\cite{deperrois2022learning, deperrois2023learning}.

\section{Preliminaries}
\label{sec_pre}

This section covers two pivotal topics essential for the subsequent sections. First, we introduce the GAN framework and its two distinct types. Our paper utilizes the GAN framework as a foundation for modeling visual hallucinations. Subsequently, we review the neurological perspective of hallucination and the brain regions implicated, drawing insights from prior research work in this field.


\subsection{Generative Adversarial Network (GAN)}
\label{sec_GAN}
\begin{figure*}[t]
	\begin{center}
	\subfloat[GAN]{%
		\includegraphics[clip,width=0.312\columnwidth,height=0.2\linewidth]{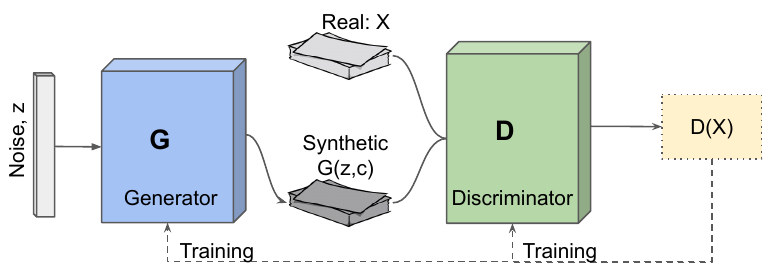}%
  \label{VGANs}
	}%
 \hfill%
	\subfloat[CGAN]{%
		\includegraphics[clip,width=0.312\columnwidth, height=0.2\linewidth]{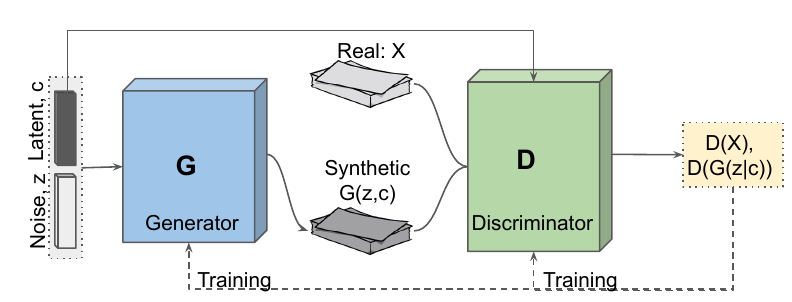}%
  \label{CGANs}
}%
 \hfill%
	\subfloat[ACGAN]{%
		\includegraphics[clip,width=0.312\columnwidth, height=0.2\linewidth]{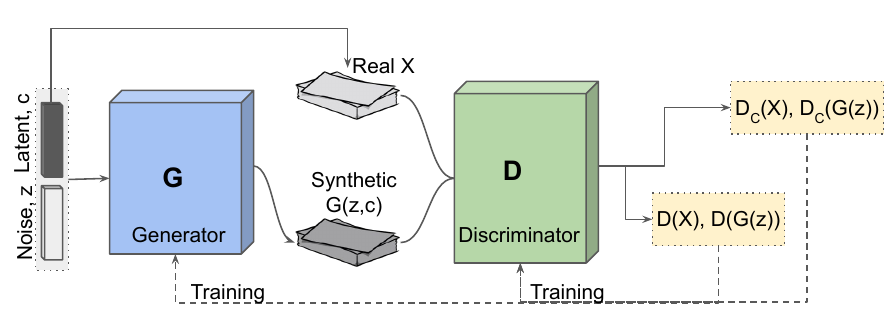}%
  \label{ACGANs}
}
\end{center}
	\caption{The overall framework of GAN architecture. GAN contains a generative network and a discriminative network. The generator generates a new image with random inputs. This generated image is sent to the discriminator alongside real images. The discriminator takes input images and classifies them into two classes: real and fake. Two types of GAN are shown in (b), and (c). Figures created by the authors.}
	\label{GANs}
\end{figure*}

A GAN is a generative model that uses
a generator and discriminator networks in an adversarial setting (Fig \ref{VGANs}). 
GANs can be used for both semi-supervised and unsupervised learning~\cite{lan2020generative}. 
The discriminator network has a training set consisting of samples drawn from the distribution $p_{\rm data}$ and learns to represent an estimate of the distribution. As a result, the discriminator network can classify the given input as real or synthetic. 
The generator network maps noise variables $z$ onto synthetic samples, according to the prior distribution of the noise variables
$P_{z}(z)$.
This way, the generator and discriminator networks contest in a two-player zero-sum min-max game. 
 The relevant objective function of a plain vanilla GAN~\cite{lan2020generative} using the Jensen-Shannon divergence metric can be written as:
\begin{eqnarray}
\label{GAN_Eq}
&&\min_G\max_D\left\{ E_{x\sim P_{\rm data}(x)}[\log D(x)]+E_{z\sim P_{z}(z)}[\log (1-D(G(z)))]\right\},
\end{eqnarray}
where $x$ represents real data samples, while $z$ represents noise. The function $G(z)$ is the generator, which takes noise samples $z$ 
 as input and transforms it into an estimated data distribution ($p_g$). On the other hand, $D(x)$ maps the input data distribution $P_{data}$ to a value in the range [0, 1]. $D(x)$ indicates the probability of a sample being real and not generated by the generator.

We utilize two variations of the basic GAN namely, Conditional GAN (CGAN)~\cite{mirza2014conditional} and Auxiliary Classifier GAN (ACGAN)~\cite{odena2017conditional}.
 to generate visual hallucinations. A CGAN is a modified version of GAN that involves providing auxiliary information to $G(x)$ during the generation process~(shown as Fig.~\ref{CGANs}). By incorporating auxiliary information during training, CGAN enables the network to distinguish between different classes of images. This allows us to instruct our model to generate images of particular objects by providing the corresponding auxiliary information. 
 The auxiliary information $c$ (known as the latent information or latent vector) and the noise vector 
 are integrated into joint hidden representations in the CGAN generator.
The data distributions are now conditioned on $c$. The loss function in a CGAN is expressed as follows:
 \begin{eqnarray}
\label{CGAN_Eq}
&&\min_G\max_D\left\{ E_{x\sim P_{\rm data}(x)}[\log D(x|c)]+E_{z\sim P_{z}(z)}[\log (1-D(G(z|c)))]\right\}.
\end{eqnarray}

ACGAN is a type of GAN that integrates class information into the GAN framework, allowing for more specific and targeted outputs (see Fig.~\ref{ACGANs}).
The ACGAN model can perform semi-supervised learning by disregarding the component of the loss occurring from class labels when a label is unavailable in the training set.
The objective functions of ACGAN are expressed as follows:
\begin{eqnarray}
\label{AUXGAN_1_Eq}
L_s&=& E_{x\sim P_{\rm data}(x)}[\log D_{adv}(x)]+E_{z\sim P_{z}(z)}[\log (1-D_{adv}(G(z)))]
\end{eqnarray}
and 
\begin{eqnarray}
\label{AUXGAN_2_Eq}
L_c&=& E_{x\sim P_{\rm data}(x)}[\log D_{cls}(c|x)]+E_{z\sim P_{z}(z)}[\log (D_{cls}(c|G(z)))]
\end{eqnarray}
The discriminator network is trained to classify~($D_{cls}$) and differentiate between real and fake images~($D_{adv}$). It aims to maximize $L_c+L_s$. On the other hand, the generator's objective is to deceive the discriminator by generating high-quality images specific to certain classes. Therefore, it is trained to maximize $L_c-L_s$.

 \begin{figure}[!t]
	\begin{center}
		\includegraphics[scale=0.27]{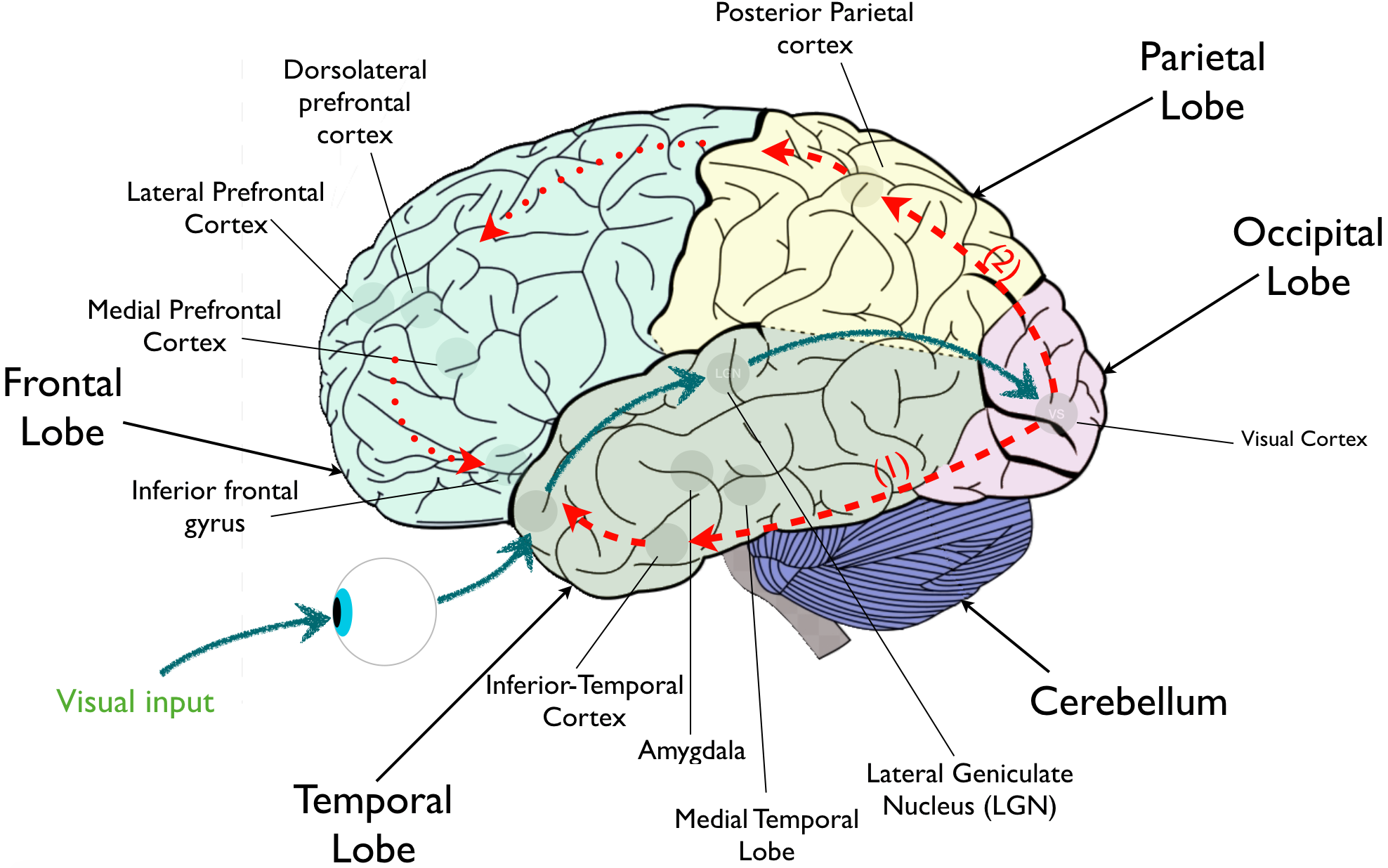}
	\end{center}
	\caption{Functional anatomy of a healthy human brain with regard to vision. Figure created by the authors.}
	\label{BA}
\end{figure}

\subsection{Hallucination}
\label{H}

In a healthy brain, when an indvidual sees an object, some areas in the brain interact with each other to perceive the object. Fig.~\ref{BA} shows the functional anatomy of a healthy human brain regarding vision. Given that vision constitutes the predominant sensory modality for a substantial portion of the human population, the allocation of visual attention is integral to advanced cognitive functions. Consequently, impairments in visual attention are often observed as a central symptom in numerous neuropsychiatric and neurological disorders~\cite{lockhofen2021neurochemistry}. 
As shown in Fig.~\ref{BA}, the information passes from the retina via the optic nerve and optic tract to the lateral geniculate nucleus (LGN) in the thalamus. The signals project from there via the optic radiation to the primary visual cortex cells, which process simple local attributes such as the orientation of lines and edges. From the primary visual cortex, information is organized as two parallel hierarchical processing streams \cite{weil2016visual}:

\begin{itemize}
\item[1)] \emph{The ventral stream}, which identifies the features of the objects and passes them from the primary visual cortex to the inferior temporal cortex.

\item[2)]\emph{The dorsal stream}, which processes spatial relations between objects and projects through the primary visual cortex to the superior temporal and parietal cortices.
\end{itemize}	

Finally, the prefrontal cortex areas (such as the inferior frontal gyrus and the medial prefrontal cortex) analyze the data received from other areas, from a real/fake point of view.

If the connectivity between any of the above-explained brain areas is disrupted, humans cannot understand the object or may perceive it falsely.
A relatively common form of memory distortion arises when individuals must discriminate items they have seen from those they have imagined (reality monitoring)~\cite{kensinger2006neural}. 
In some neuro-diseases, individuals cannot discriminate whether an item was imagined or perceived from the external environment. In this regard, hallucination is defined as the unpredictable experience of perceptions without corresponding sources in the external world~\cite{bilder2013neuroscience}.

Now, in order to model the interactions between different brain areas with regard to hallucinations, we look into the known or suggested causes for the incidence of hallucinations. In particular, some studies show that hyperdopaminergic activity in the hippocampus makes hallucinations in schizophrenia~\cite{silbersweig1995functional, moustafa2016cognitive}. 
Also, a grey matter volume reduction is seen in  Parkinson's disease patients with visual hallucinations involving occipito-parietal areas associated with visual functions~\cite{ramirez2007cerebral}.
The hippocampal region dysfunction and abnormalities in GABA ($\gamma$-Aminobutyric Acid) and dopamine function are seen to have a role in causing this disease~\cite{moustafa2010neural}. Abnormal cortical dopamine D1/D2 activation ratio may be related to altered GABA and glutamate transmission~\cite{csaszar2019possible}. Moreover, neurotransmitters such as norepinephrine, acetylcholine, and dopamine are thought to impact different aspects of attention~\cite{lindsay2020attention, lockhofen2021neurochemistry}. Norepinephrine is associated with alertness, acetylcholine with orienting to important information, and dopamine with executive control of attention~\cite{lindsay2020attention, lockhofen2021neurochemistry}.

In order to model hallucination, we consider the areas of the brain involved in hallucination, according to the previous relevant studies \cite{weil2016visual,moustafa2016cognitive}.
Visual hallucinations in Parkinson's disease are caused by overactivity of the Default Mode Network (DMN) and Ventral Attention Network (VAN) and under-activity of the Dorsal Attention Network (DAN)~\cite{weil2016visual}. VAN  mediates the switch between DAN and DMN. The overactivity of DMN and VAN reinforces false images, which DAN fails to check when underactive~\cite{weil2016visual}.
Moreover, on functional neuroimaging studies, patients with visual hallucinations showed decreased cerebral activation in occipital, parietal, and temporoparietal regions, and increased frontal activation in the region of frontal eye fields~\cite{diederich2009hallucinations}.
%
\begin{figure}[t]
	\begin{center}
		\includegraphics[scale=0.41]{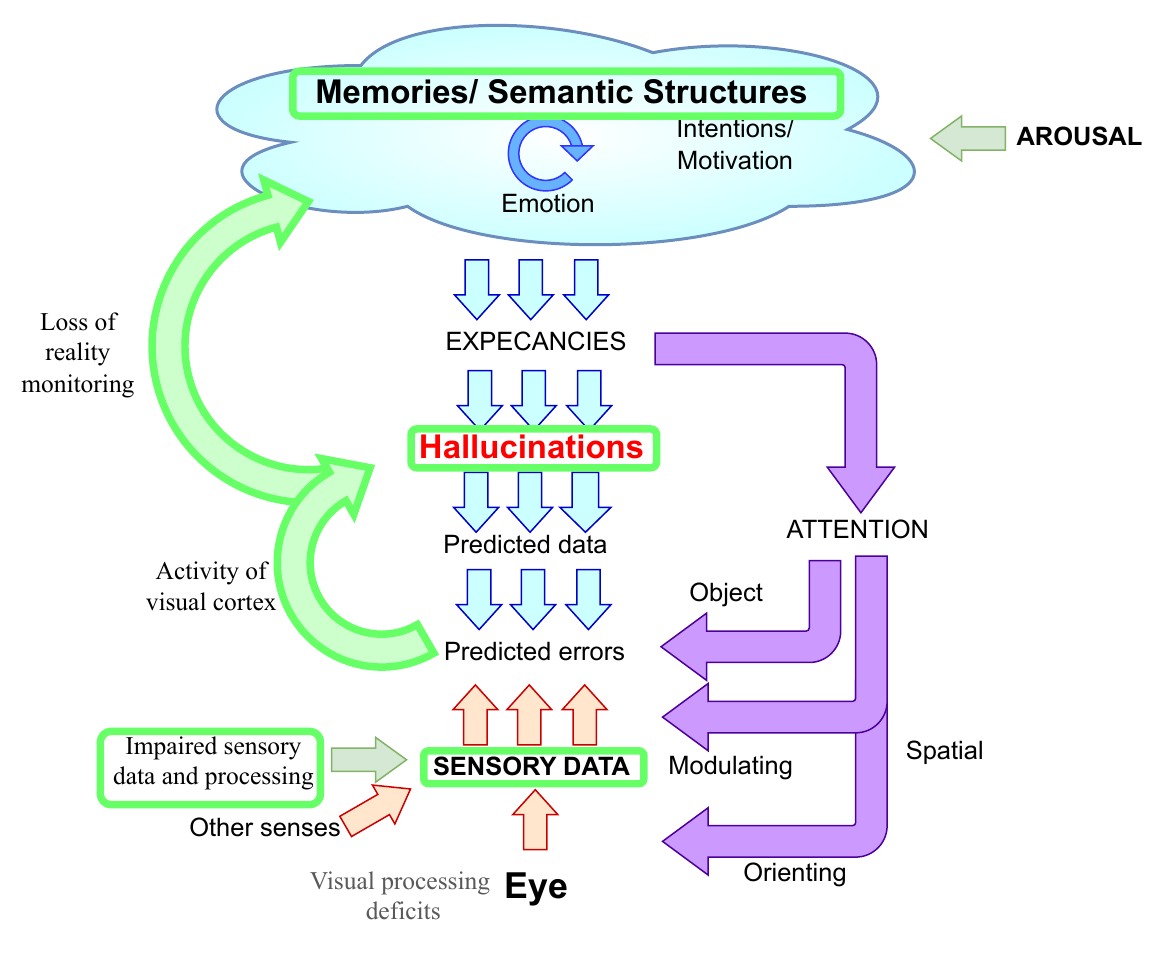}
	\end{center}
	\caption{Framework of visual perception. We showed details of our idea and its relationship to the hallucination framework (Adapted from~\cite{collerton2023understanding}) based on our findings, highlighted in bold green box. Figure created by the authors using \href{draw.io}{draw.io} software.}
	\label{hallu_d}
\end{figure}
It is important to note that brain connections are not static but rather dynamic, as they change all the time. Considering the aforementioned areas involved in hallucinations, and the effect of neurotransmitters in the connectivity between different areas of the brain, one can conclude that an imbalance between dopamine, acetylcholine, and other neurotransmitters is involved in the pathogenesis of visual hallucinations.

As shown in Fig.~\ref{hallu_d}, disruptions between visual areas and cognitive areas could potentially occur at several levels. Inspired by those mentioned above, a GAN-based model provides a better visual representation of brain functions in a hallucination state than diffusion and autoencoder models because its networks interact with each other through an adversarial mechanism.   In Section~\ref{HM}, we present a theoretical GAN-based model for hallucinations, which highlights the functional importance of brain areas,  their connections, and neurotransmitters.

\begin{table*}[t]
\renewcommand{\arraystretch}{1.5}
	\caption{GAN and Brain with hallucination}
	\label{matching}
	\centering
{\small
	\begin{tabular}{
			|>{\columncolor[HTML]{C0C0C0}}c|c|c|}
		\hline
		\multicolumn{1}{|c|}{\cellcolor[HTML]{9B9B9B}}                            & \multicolumn{2}{|c|}{\cellcolor[HTML]{C0C0C0}}                                                                                                                   \\ \cline{2-3}
		\multicolumn{1}{|c|}{\multirow{-2}{*}{\cellcolor[HTML]{9B9B9B}\diagbox[width=\dimexpr \textwidth/8+3\tabcolsep\relax, height=1cm]{ Attribute}{Models}}} & \multicolumn{1}{|c|}{\multirow{-2}{*}{\cellcolor[HTML]{C0C0C0} Brain with Hallucination}}  & \multicolumn{1}{|c|}{\multirow{-2}{*}{\cellcolor[HTML]{C0C0C0} Hallu-GAN}} \\ \hline
		Generator                                                                &  Occipital lobe, Visual cortex, and Parietal area                                                          &    Artificial Neural network                                        \\ \hline
		Discriminator                                                            & Prefrontal cortex and Inferior frontal gyrus                                                       &      Artificial Neural network                                               \\ \hline
		Input of Discriminator                                                   & Brain Signal                                                                  &      Images                                                                              \\ \hline
		Output of Discriminator                                                  &   Imagination or Real                                                         & Hallucination or Non-Hallucination                                                                          \\ \hline
		Input of Generator                                                       &    Nothing or Noise and Environment’s Image                                                             & Noise and Environment’s Image                                                                              \\ \hline
		Output of Generator                                                      &  Imagination                                                                &       Fake Image                                                                                  \\ \hline
		Neuron                                                                   &        Interneurons and pyramidal neurons                                                       &    Artificial  Neuron                                                                                  \\ \hline
	\end{tabular}
}
\end{table*}

\section{Hallu-GAN: The Proposed Approach to Modeling Hallucination}
\label{HM}

In this section, we present our proposed GAN model for hallucination. 
Different types of retrieved information (e.g., perceptual detail, information about cognitive operations) are typically to determine whether an item was imagined or perceived from the external environment. As explained in the previous section, a breakdown in the connectivity of neural networks and dysfunction of some brain areas are known to result in visual hallucinations.
Indeed, some brain areas, especially the occipital lobe, the visual cortex, and the parietal area, change their mechanisms.
Specifically, they process imperfect visual input data and send output to other areas. This process somehow mimics the role of the generator in a GAN, trying to change the visual input data in order to deceive the other areas that were responsible for the perception between reality and imagination (resembling the discriminator of a GAN).
In particular,
some cortical areas, especially the prefrontal cortex, and inferior frontal gyrus process the input to determine whether an item was imagined or perceived.
As mentioned in Section~\ref{H}, the perturbations in some neurotransmitters, especially dopamine, impact the functionality of these areas. As a result, these areas cannot truly classify the input to determine whether an item was imagined or perceived. This imperfect functionality thus initiates a contest between the distinguishing region and the falsifying region which functions in an adversarial setup. Putting the two aforementioned sides together,  the adversarial interaction between the mentioned areas of the brain can be viewed as a GAN.
Table~\ref{matching} summarizes the correspondence between the elements of the hallucinating brain and their counterparts within the relevant GAN model.

\begin{figure*}[!t]
	\begin{center}
		\includegraphics[scale=0.22]{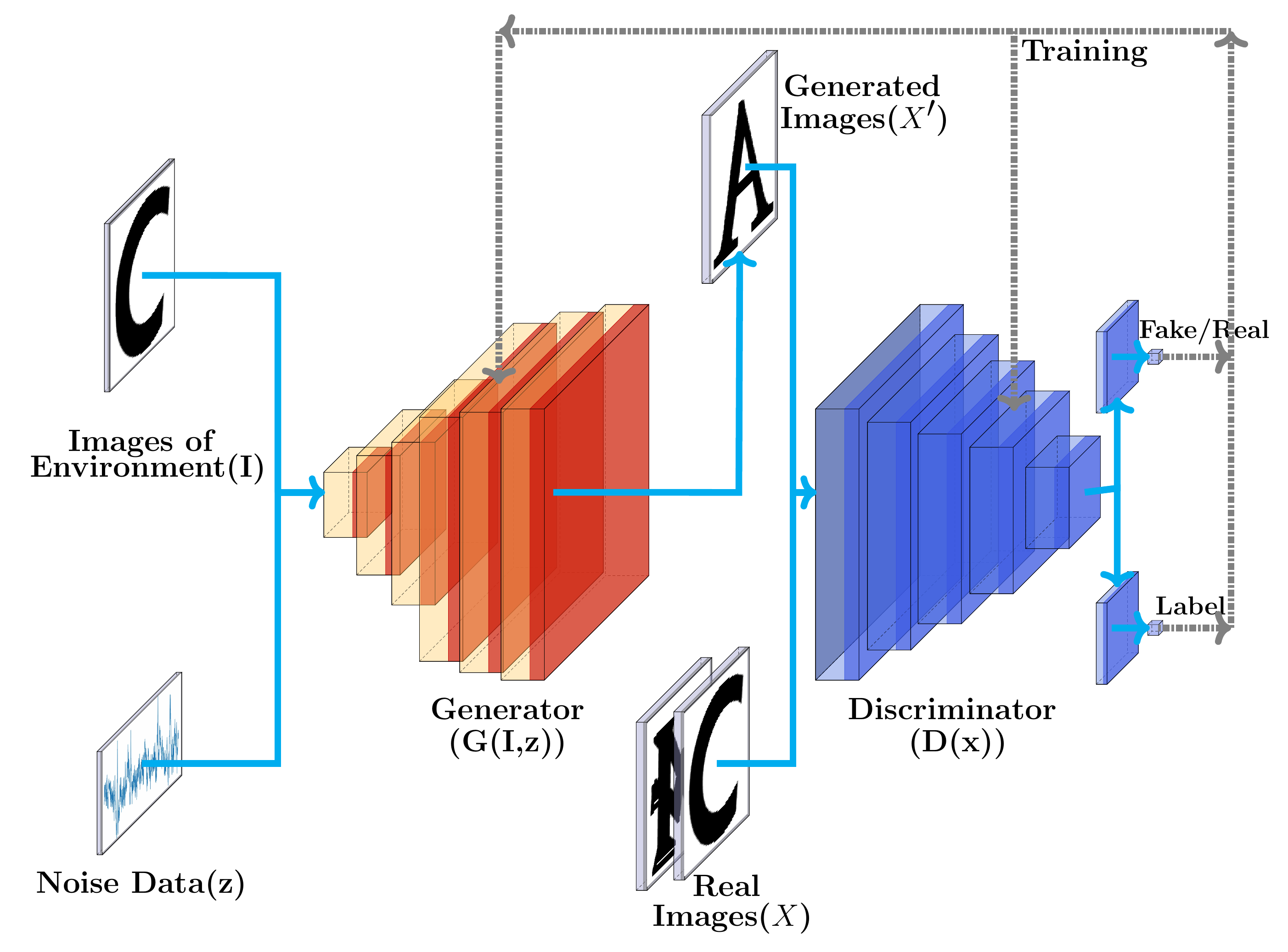}
	\end{center}
	\caption{Illustration of Hallu-GAN model. Hallu-GAN model has two neural networks (Generator and Discriminator). 
    The generator input is just the environment's image with input noises. The discriminator has two tasks. The discriminator distinguishes either between real and fake output or between non-hallucination and hallucination output. Figure created by the authors.}
	\label{Aux_GAN_1}
\end{figure*}

In our proposed model, the generator takes both an environmental image and a random vector as inputs. Based on the training state (healthy and hallucinatory states), it generates an image.  The discriminator must distinguish between real and fake outputs, and also between non-hallucinatory and hallucinatory outputs. 
Thus, how can the discriminator network accomplish these two types of tasks?
Taking into account the structure of the brain, in the following, we propose a \textbf{Hallu}cinatory Auxiliary Classifier Conditional GAN (Hallu-GAN). Hallu-GAN combines a CGAN and an ACGAN, as shown in Fig.~\ref{Aux_GAN_1}. Incorporating convolutional neural networks (CNNs) into our proposed model would be advantageous as they were designed based on the principles of biological vision~\cite{lindsay2021convolutional}.

The generator ($G$) of Hallu-GAN is identical to the generator in a CGAN. $G$ takes as input an environment’s image and 
a noise vector ($z$) to generate images. The discriminator ($D$) used in our approach is the same as the one employed in ACGAN.
$D$ produces two outputs: the first output is the probability of input data being real or fake, and the second output is the estimated conditional probability of the class label given the input data.
There are two parts to the objective function: the log-likelihood of the correct source, $L_S$, and the log-likelihood of the correct class, $L_C$ given by
\begin{eqnarray}
\label{hallGAN_1_Eq}
L_s&=& E_{x\sim P_{\rm data}(x)}[\log D_{adv}(x)]+E_{z\sim P_{z}(z)}[\log (1-D_{adv}(G(z,I)))]
\end{eqnarray}
and
\begin{eqnarray}
\label{hallGAN_2_Eq}
L_c&=& E_{x\sim P_{\rm data}(x)}[\log D_{cls}(c|x)]+E_{z\sim P_{z}(z)}[\log (D_{cls}(c|G(z,I)))]
\end{eqnarray}
$G$ in this model uses just the environment’s image and the noise ($z$) to generate output images. According to two states (hallucinatory and healthy states), $D$ in this model is trained to maximize $L_S + L_C$, while $G$ is also trained to maximize $L_C - L_S$. However, the generator is only trained based on one of the two states, namely the hallucinatory or healthy state.

\begin{figure*}[!t]
	\begin{center}
		\includegraphics[scale=0.2]{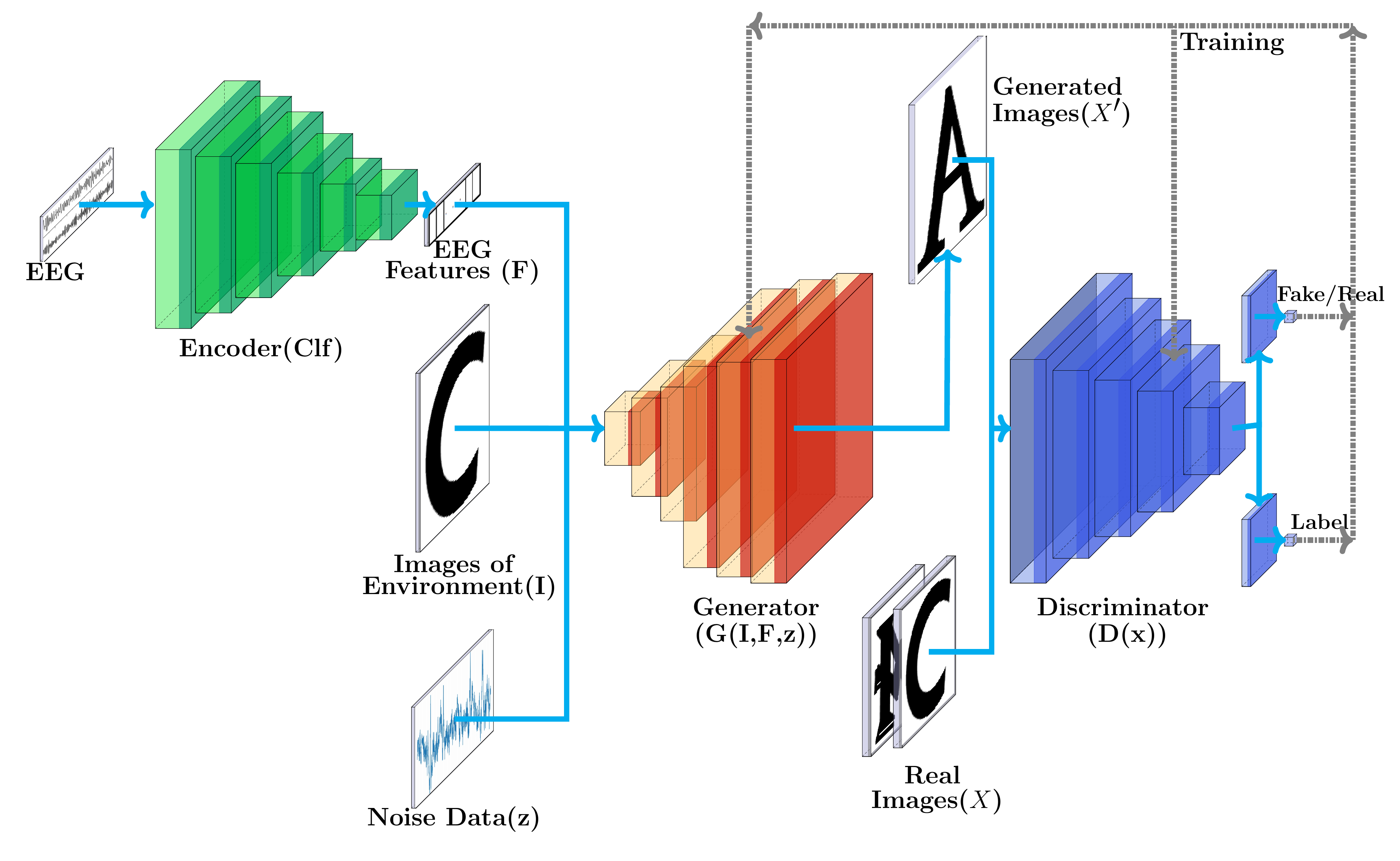}
	\end{center}
	\caption{Illustration of Hallu-GAN$^+$ model.  Hallu-GAN$^+$ model has three networks (Encoder/Classifier, Generator, and Discriminator). Most layers of  Hallu-GAN$^+$ model are convolutional networks. The task of the Encoder/Classifier is extracting features from EEG data.  The generator has three inputs (EEG data, Image of the environment, and input noises). The discriminator has two tasks. The discriminator distinguishes either between real and fake output or between non-hallucination and hallucination output. Figure created by the authors.}
	\label{Aux_GAN_2}
\end{figure*}

The functionality of other brain regions influences the operation of the visual system during hallucinations~\cite{diederich2009hallucinations}. In particular, memory disturbances play a critical role in visual hallucinations~\cite{barnes2011visual, brebion2007visual, brebion2012source, brebion2019clinical}. To model the impact of these other regions on the visual system, we modify the model's inputs by adding a new element to determine the state of the brain in the generator. To arrive at a mechanism similar to the brain, we take  EEG/fMRI data as input into the generator network. The generator ($G$) network thus receives the environment's image, noise, and EEG/fMRI features and produces a new image.  This model is called Hallu-GAN$^+$ and its architecture is shown in Fig.~\ref{Aux_GAN_2}. In this model, the discriminator ($D$) network is a single component that does both of the required supervised and unsupervised discrimination tasks.
There are two parts to the objective function: the log-likelihood of the correct source, $L_S$ given by
\begin{eqnarray}
\label{hallGAN+_1_Eq}
L_s&=& E_{x\sim P_{\rm data}(x)}[\log D_{adv}(x)]+E_{z\sim P_{z}(z)}[\log (1-D_{adv}(G(z,I,F)))]
\end{eqnarray}
and the log-likelihood of the correct class, $L_c$ given by
\begin{eqnarray}
\label{hallGAN+_2_Eq}
L_c&=& E_{x\sim P_{\rm data}(x)}[\log D_{cls}(c|x)]+E_{z\sim P_{z}(z)}[\log (D_{cls}(c|G(z,I,F)))]
\end{eqnarray}

$D$ is trained to maximize $L_S + L_C$, while $G$ is trained to maximize $L_C - L_S$. Hallu-GAN$^+$ learns a representation for $z$ that is independent of the class label.
This would allow the model to assesses if an environment is calm or chaotic.
In terms of training our models, the generator of the first proposed model takes  the environmental image and noise as inputs to produce an output image. Following that, the environmental image, noise, and EEG/fMRI data are inputs to the generator in the second proposed model, which produces an output image. Lastly, both our proposed models use real and artificially generated images to train their discriminators to recognize hallucinations. 

The generative adversarial perspective, unlike  Bayesian models, suggests a broad hypothesis about the origin of hallucination content (via an abnormal generator) similar to that of delusion~\cite{gershman2019generative}.


\section{HalluGAN Performance Evaluation}
\label{Sim_model}
To illustrate the capabilites of the proposed models, we apply the first learning algorithm presented in Section~\ref{HM}.  In the next  subsections, we will  explain how we create a (hypothetical) experimental hallucination dataset from existing real data. We will then explain the evaluation metrics in Section~\ref{metrics}. We will finally provide the results of the trained models in Section~\ref{Results}.

\subsection{Dataset Construction}
\label{Data_sec}
\begin{figure}[!t]
	\begin{center}
	\subfloat[Examples of  the dataset with three different subsets: Digits, Characters, and Objects.]{%
		\includegraphics[clip,width=0.48\columnwidth,height=0.39\linewidth]{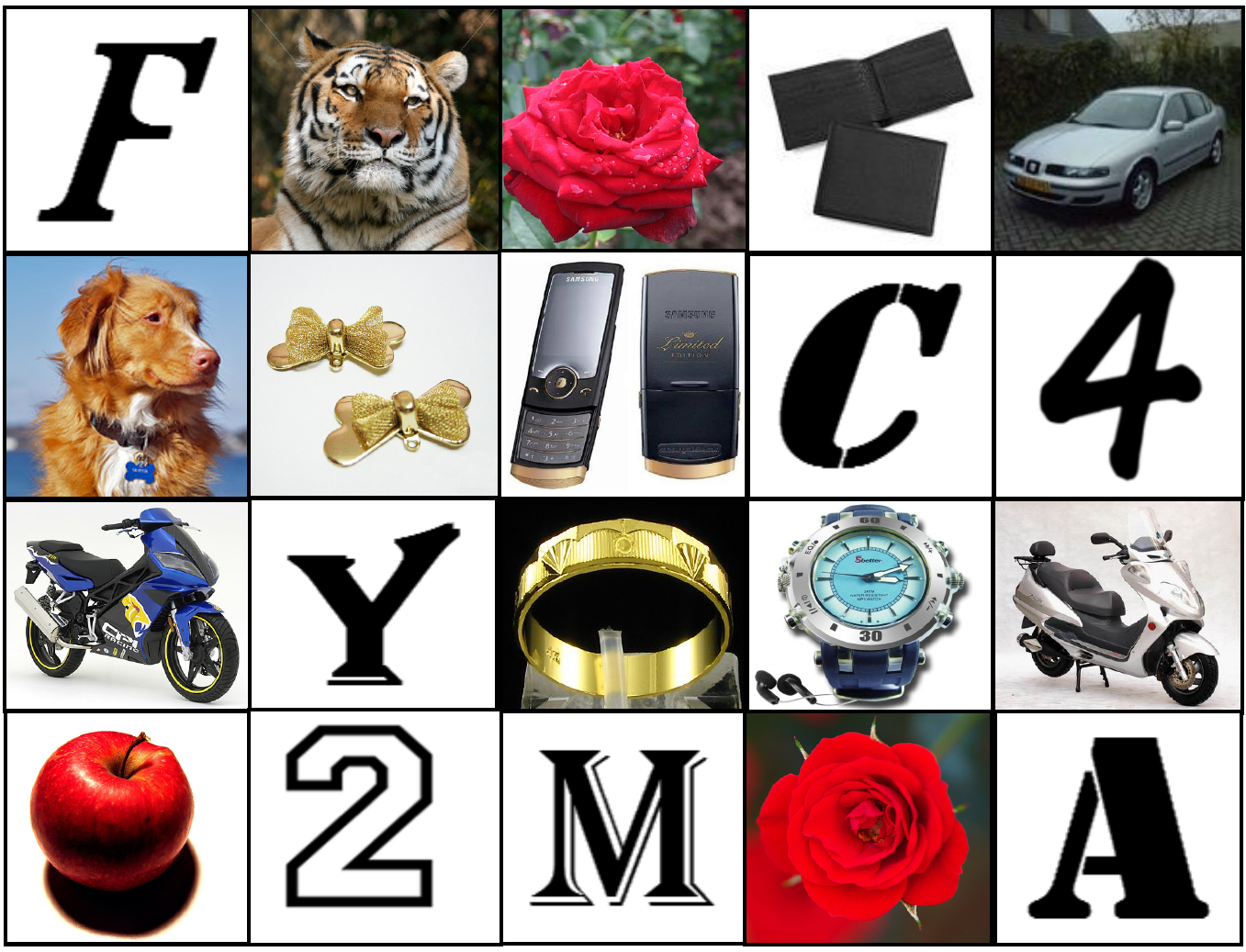}%
        \label{Dataset_thviz}
	}%
 \hfill%
	\subfloat[Examples of training images  and  test images on the dataset.This dataset has two classes ('A' and 'C').]{%
		\includegraphics[clip,width=0.48\columnwidth, height=0.39\linewidth]{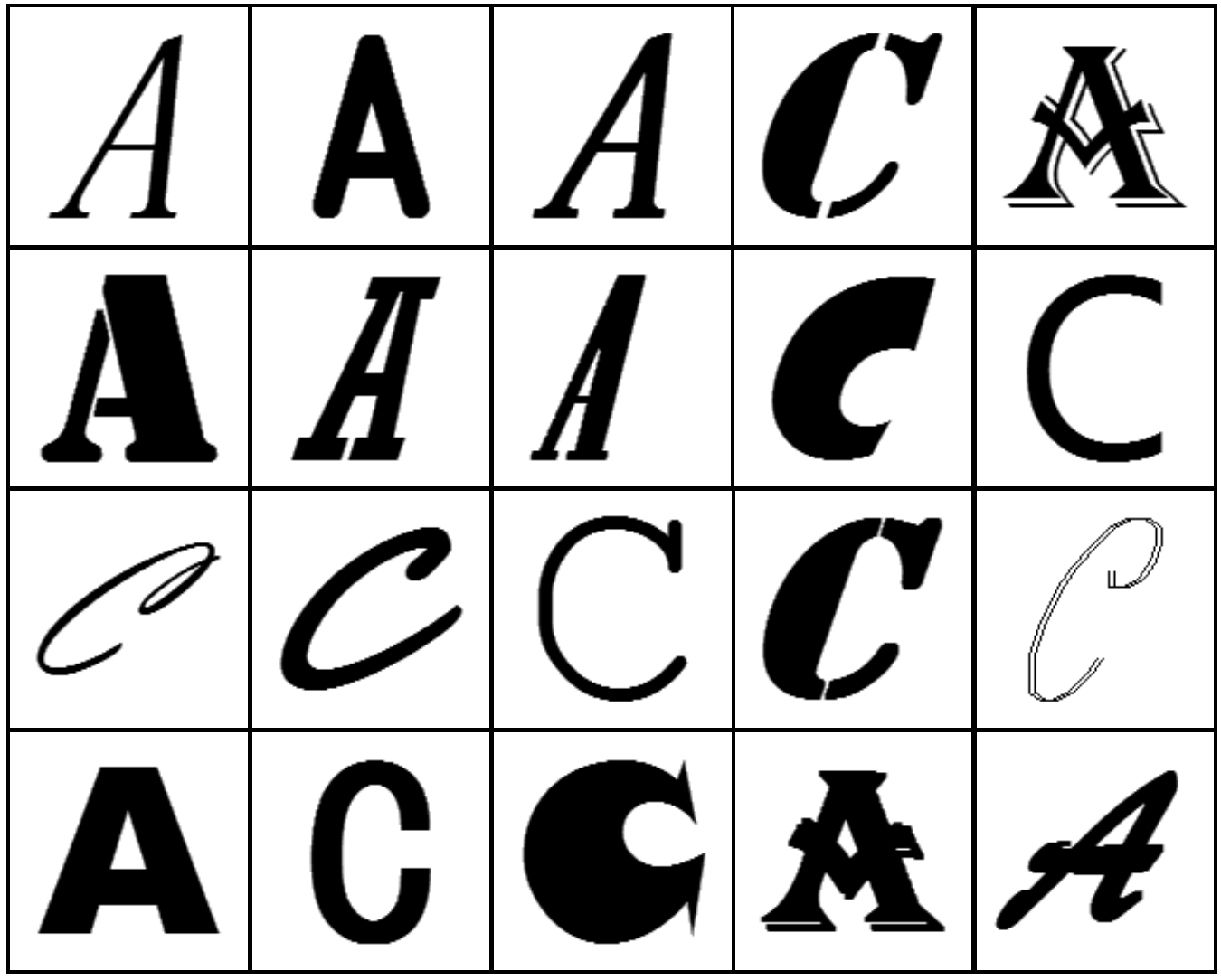}%
  \label{Dataset_ca}
}
\end{center}
	\caption{Examples of the used datasets~\cite{kumar2018envisioned, tirupattur2018thoughtviz}. The figures are under public dataset~\cite{kumar2018envisioned}.}
	\label{Dataset}
\end{figure}

To train and test the Hallu-GAN models, first, we construct a new dataset based on the avialble real data of ~\cite{kumar2018envisioned, tirupattur2018thoughtviz}. The actual dataset contains EEG recordings of participants observing images from three different subsets: Digits, Characters, and Objects~\cite{kumar2018envisioned, tirupattur2018thoughtviz} (shown in Fig.~\ref{Dataset_thviz}). 
In our experiments, we use only two classes ('A' and 'C') yielding a dataset containing a total of 2032 images (from both classes) with their corresponding EEG data. Fig.~\ref{Dataset_ca} shows some exemplars of the constructed dataset. The images of the two classes, and their corresponding EEG signals are used as follows to provide an hypothetical hallucinatory framework. 
An image of the 'C' character is considered as the input image representing  environ's image. 
Therefore, the EEG data corresponding to the 'A' characters are viewed as hallucinatory data while the the EEG data of the 'C' characters represent the  healthy non-hallucinatory state.

In the next step, we use the EEG data of an individual experiencing Charles Bonnet's syndrome including both his resting state and his hallucination state, from the study in \cite{piarulli2021high} which has been approved by the Ethics Committee of
the Faculty of Medicine of the University of Liège.

Charles Bonnet syndrome refers to the occurrence of hallucinations, where one perceives things that are not present. This condition is commonly observed in individuals who have experienced significant loss of vision. Since the individual is blind, we utilize a black image as the input of the  generator and the resting state set. While the indivual's hallucinations have been reported to range from simple flashes or colored backgrounds to more complex scenes with the appearance of faces, objects, people, or landscapes ~\cite{piarulli2021high}, for the sake of modeling convenience here, we once again employ the 'C' set of images to represent the hallucination state in general.

\subsection{Evaluation Metrics}
\label{metrics}
We use the following metrics to evaluate the performance of the proposed models. The first metric we use is the Inception Score (IS), which is a mathematical metric used to assess the quality of images generated by a GAN. 
IS is expressed as follows: 
%
\begin{equation}
\label{IS}
\small IS=
exp(E_{x\sim P_{\rm g}D_{KL}}[(p(y|x)||p(y)]),
\end{equation}
where $D_{KL}$ is the KL divergence between $p(y|x)$ and $p(x)$, $x$ is the image produced by the generator, and $y$ is the expected label of $x$. 
Another metric is Mutual Information (MI) which is a measure of the mutual dependence between two random variables in probability theory. It quantifies the amount of information obtained about one random variable by observing the other. High mutual information indicates a significant reduction in uncertainty, while low mutual information indicates a minimal reduction. If the mutual information between two random variables is zero, it means that the variables are independent of each other. MI is  expressed as follows:
\begin{eqnarray}
\label{MI_Eq}
MI(X;Y)&=& D_{KL}(P_{XY}(x,y)||P_X(x)P_Y(y)),
\end{eqnarray}
where a pair of random variables $(X, Y)$ have values distributed throughout the space $X \times Y$, $P_{(X, Y)}$ is their joint distribution and $P_{Y}$ and $P_{X}$ are their marginal distributions.

\subsection{Results}
\label{Results}
Here, we examine whether our generative model trained to create and reconstruct brand-new objects in images can exhibit representations resembling hallucinatory processes in the ventral visual pathway.
We experiment with three architectures in order to test the proposed models.  
The experiments are implemented on Python 3.8 by the Pytorch package.

We trained our adversarial nets on a dataset as explained in Section~\ref{Data_sec}. The generator nets used a mixture of Relu activations and Tanh activations, while the discriminator net was used to max out activations.
As described in Section~\ref{metrics}, in order to assess the generator's performance we use three metrics: accuracy/loss, MI, and IS.

\subsubsection{\textbf{Hallu-GAN} }

In the first experiment, we use our first model (shown in Fig.~\ref{Aux_GAN_1}). We implement two Hallu-GAN models with the same architecture.
The first Hallu-GAN model is to represent a healthy person, while the second Hallu-GAN model represents a diseased person. We train each Hallu-GAN model using only one set of 'C' and 'A' images,  scaling all training images to a resolution of $64\times 64$ pixels. This gives 1000 training images in total for each model.

\begin{figure}[!t]
	\begin{center}
		\includegraphics[clip,width=0.48\columnwidth, height=0.39\linewidth]{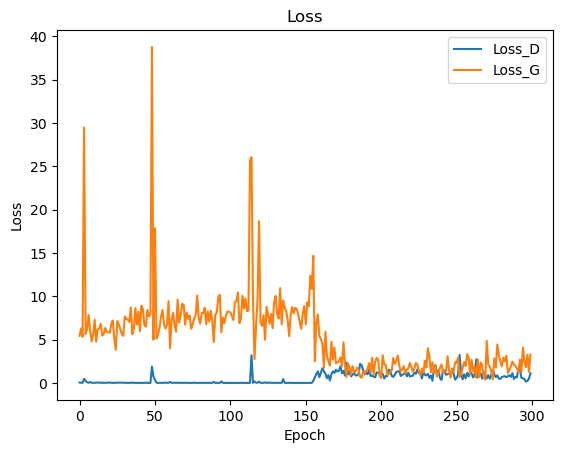}%
\end{center}
	\caption{Loss curves of Hallu-GAN training during the hallucinating state. The losses of the generator (G) and discriminator (D) stabilize after 300 epochs of training, reflecting the convergence of the proposed model.}
	\label{loss_hallu}
\end{figure}

\begin{figure}[!t]
	\begin{center}
	\subfloat[Hallucination state while the 'C' is input.]{%
		\includegraphics[clip,width=0.48\columnwidth,height=0.39\linewidth]{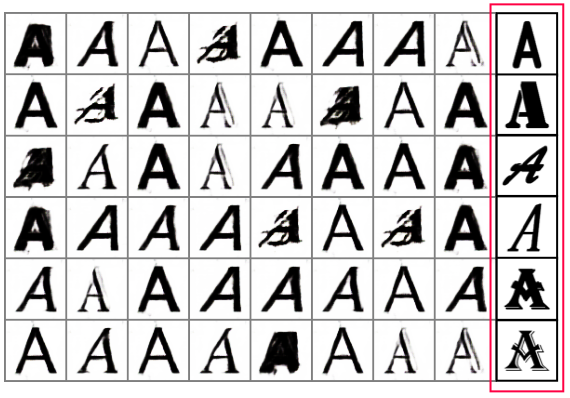}%
        \label{c_a_images}
	}%
 \hfill%
	\subfloat[Healthy state for which we expect the output 'C'.]{%
		\includegraphics[clip,width=0.48\columnwidth, height=0.39\linewidth]{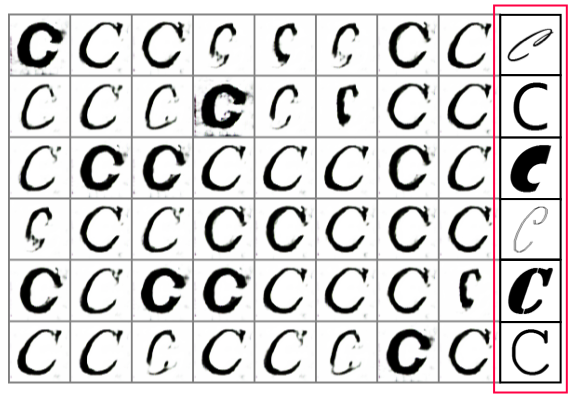}%
  \label{c_c_images}
}
\end{center}
	\caption{The results of Hallu-GAN after training. Given the environmental input character 'C', the model generates a corresponding output. (a) The results of Hallu-GAN depict a hallucinating visual system. (b) The results of Hallu-GAN illustrate a healthy visual system. It is important to note that columns 1 through 8 contain images generated by our model, while only the last column features randomly selected images from the training set. The figures of the last column are under public dataset~\cite{kumar2018envisioned}.}
	\label{hallugan}
\end{figure}

\textcolor{black}{The Hallu-GAN model converges in two states (healthy and hallucinating), and the loss associated with its networks in the training process decreases. For example, Fig.~\ref{loss_hallu} shows the loss during the training of the Hallu-GAN model in a hallucinating state. The  IS metric for Hallu-GAN is $2.71$.  Fig.~\ref{hallugan}  displays the images created using our suggested model for the hallucinating and healthy brain. 
Fig.~\ref{c_a_images} shows the output of the Hallu-GAN model in the hallucinating state and Fig.~\ref{c_c_images} shows the output of the Hallu-GAN model in the healthy state. These results demonstrate that the model can mimic the behavior of both the hallucinating and healthy brain. Now one key question here is what differences are there between the generators in the two states that has led to such distinct functionalities.}

The generator layers in the Hallu-GAN models for subjects who are hallucinating and those who are not can be compared. 
As pointed out earlier, MI is a statistical metric that can be used to examine how the layers of different models differ from one another.  
It is evident from Table~\ref{weight_correlation}  that there is a declining trend in mutual information among the generator's layers. 
Compared to the initial layer of generators, the latter layers of generators are less dependent.
So, the last layers of generators are far more effective in producing a fake image that resembles the last layers of the visual system.

\begin{table*}[!t]
\renewcommand{\arraystretch}{1.25}
\caption{Correlation of layers of the generators in two states, i.e., hallucinatory and healthy state.}
\centering
\begin{tabular}{|c|c|c|c|c|c|c|}
\hline
                   & Layer 1 & Layer 2  & Layer 3 & Layer 4 & Layer 5 & Layer 6 \\ \hline
Mutual Information         & 16.167                                                        & 15.77                                                        & 14.507                                                        & 13.16                                                      & 11.78                                                         & 8.03                                                           \\ \hline
\end{tabular}
\label{weight_correlation}
\end{table*}

\subsubsection{\textbf{Hallu-GAN$^+$}}
\label{Hallu-GAN+}
In the next experiment, we utilize our second model (shown as Fig.~\ref{Aux_GAN_2}). We train the Hallu-GAN$^+$ model using image letters 'A' and 'C', scaling all training images to a resolution of \texorpdfstring{$64\times 64$}{64*64} pixels and EEG data. There are 2000 training images in total. The generator nets used a mixture of Relu activations and Tanh activations, while the discriminator net was used to max out activations.

We incorporate an adjustable parameter ($\alpha$) between the classifier loss and discriminator loss for the generator loss ($(1-\alpha)L_C+\alpha (-L_S)$), in accordance with equation~\ref{hallGAN+_1_Eq} and equation~\ref{hallGAN+_2_Eq} to learn the model better. Since the classifier does not need to learn using the image produced by the generator in the initial stage of training, we modify the parameter alpha during the training model.  In the training process, as the generator improves and $\alpha$ increases, the classifier can observe the generator's output. The loss is calculated based on the classifier's prediction and the classifier's weights are adjusted accordingly.

\begin{figure}[!t]
	\begin{center}
    \subfloat[Output of Hallu-GAN$^+$ that the 'C' is input.]{%
		\includegraphics[clip,width=0.6\columnwidth, height=0.37\linewidth]{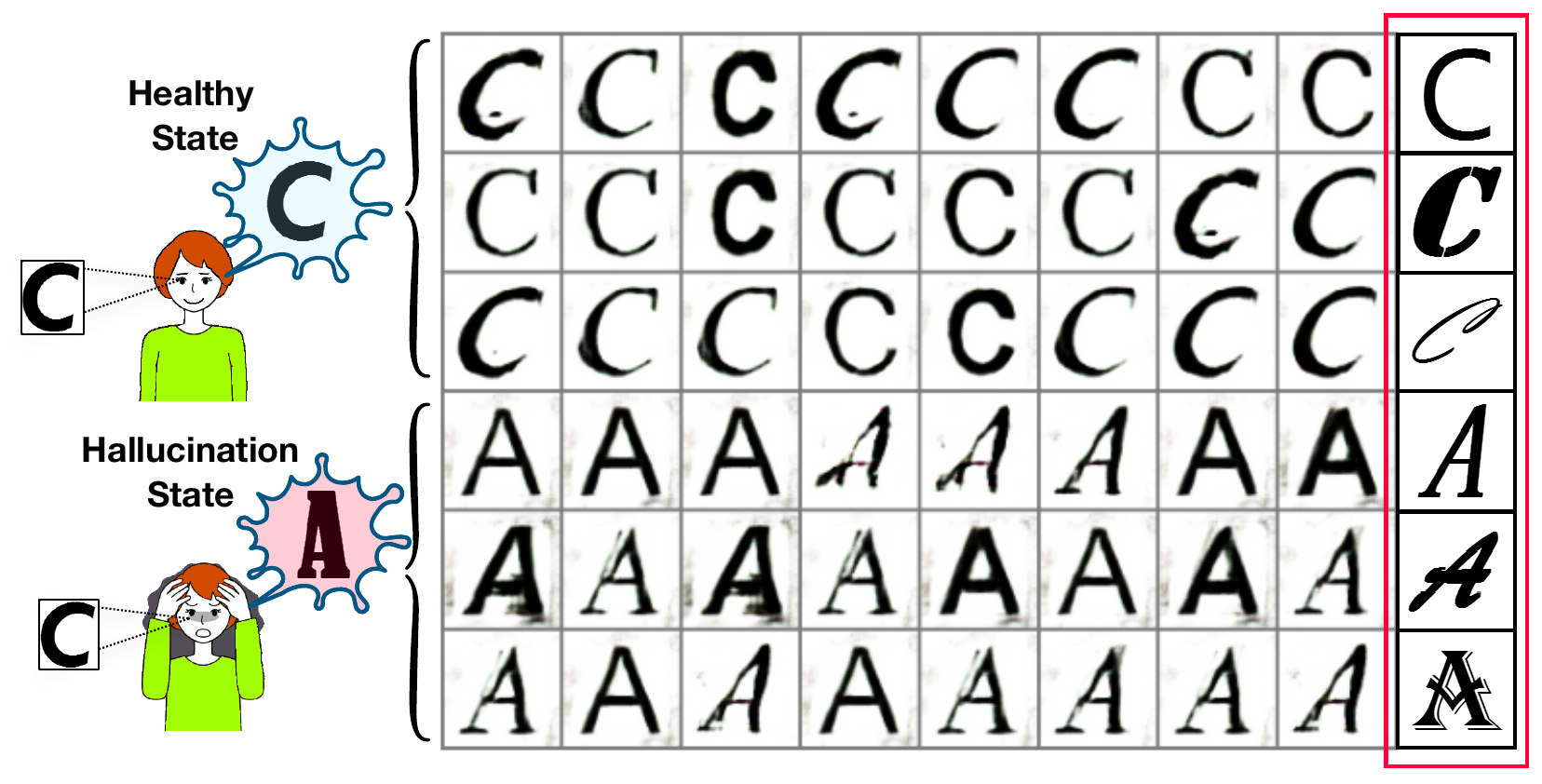}%
  \label{c_c_a}
}%
\hfill%
	\subfloat[]{%
		\includegraphics[clip,width=0.38\columnwidth,height=0.33\linewidth]{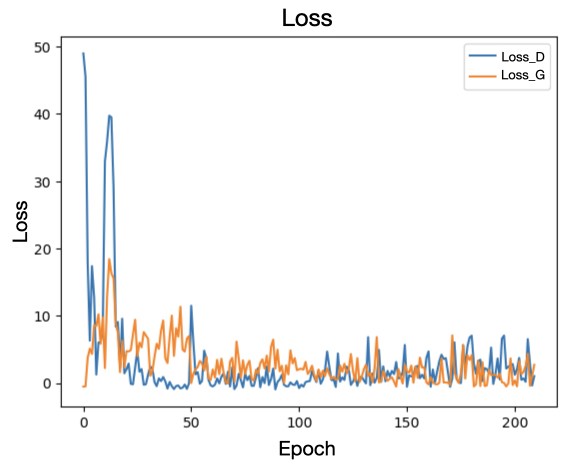}%
        \label{loss_A_C}
	}%
\end{center}
	\caption{Results of Hallu-GAN$^+$. The environmental image as the generator input is  the character 'C'. (a) Result of the Hallu-GAN$^+$  model. The first three rows of the results correspond to a hallucinating visual system.  The last three rows of results illustrate a healthy visual system. It is important to note that columns 1 through 8 contain images were generated by our model, while only the last column features randomly selected images from the training set. The figures of the last column are under public dataset~\cite{kumar2018envisioned}. (b) Loss curves of  Hallu-GAN$^+$ during training with the dataset~\cite{kumar2018envisioned}. The losses of the generator (G) and discriminator (D) stabilize after 220 epochs of training, reflecting the convergence of the proposed model.}
	\label{c-a_c}
\end{figure}

Fig.~\ref{loss_A_C} shows the loss during the training of the Hallu-GAN$^+$ model. Fig.~\ref{c_c_a} displays the images created using our suggested model for the hallucinating brain.
The last column in Fig.~\ref{c_c_a} displays a sample of randomly chosen photographs from the dataset, and the subsequent columns display the images produced by our model. The IS metric for Hallu-GAN$^+$ is 2.34.

The Hallu-GAN$^+$ model generates synthetic images that have the same distribution as  the real images  based on  healthy EEG data. Moreover, the Hallu-GAN$^+$  model produces synthetic images that closely mimic the characteristics of hallucinating images, according to supposed hallucinating EEG Data.
Hallu-GAN$^+$ is therefore capable of visualizing the hallucination process in the visual system.

The input element of the generator in Hallu-GAN$^+$ that represents the brain's current state may be a combination of the remaining senses or impairment memories~\cite{laureys2015neurology, alkam2016modeling}. 
When other areas of the brain are not functioning properly, it can disrupt brain states. This can cause abnormal behavior in certain brain regions, which can then send disruptive signals to the visual cortex. As a result, the visual system may be affected and not function properly. For future work, we can extend the model to analyze the impact of memory and the functioning of the hippocampal regions on the visual system.

\subsubsection{\textbf{Hallu-GAN$^+$ with EEG of a Patient with Charles Bonnet Syndrome}}

In the third experiment, we use our second model~(shown as Fig.~\ref{Aux_GAN_2}). We train the Hallu-GAN$^+$ model using only images of 'C', and EEG data. The training images are all scaled to a resolution of \texorpdfstring{$64\times 64$}{} pixels. This gives 2000 training images in total and the test accuracy of trained classifiers on EEG data is 88 percent. The EEG data is related to a person suffering from Charles Bonnet  syndrome~\cite{piarulli2021high}. Informed consent was obtained from the subject involved in the study. 
The generator nets used a mixture of Relu activations and Tanh activations, while the discriminator net was used to max out activations.

\begin{figure}[!t]
	\begin{center}
	\subfloat[Output of Hallu-GAN$^+$ that the black image is input.]{%
		\includegraphics[clip,width=0.6\columnwidth, height=0.37\linewidth]{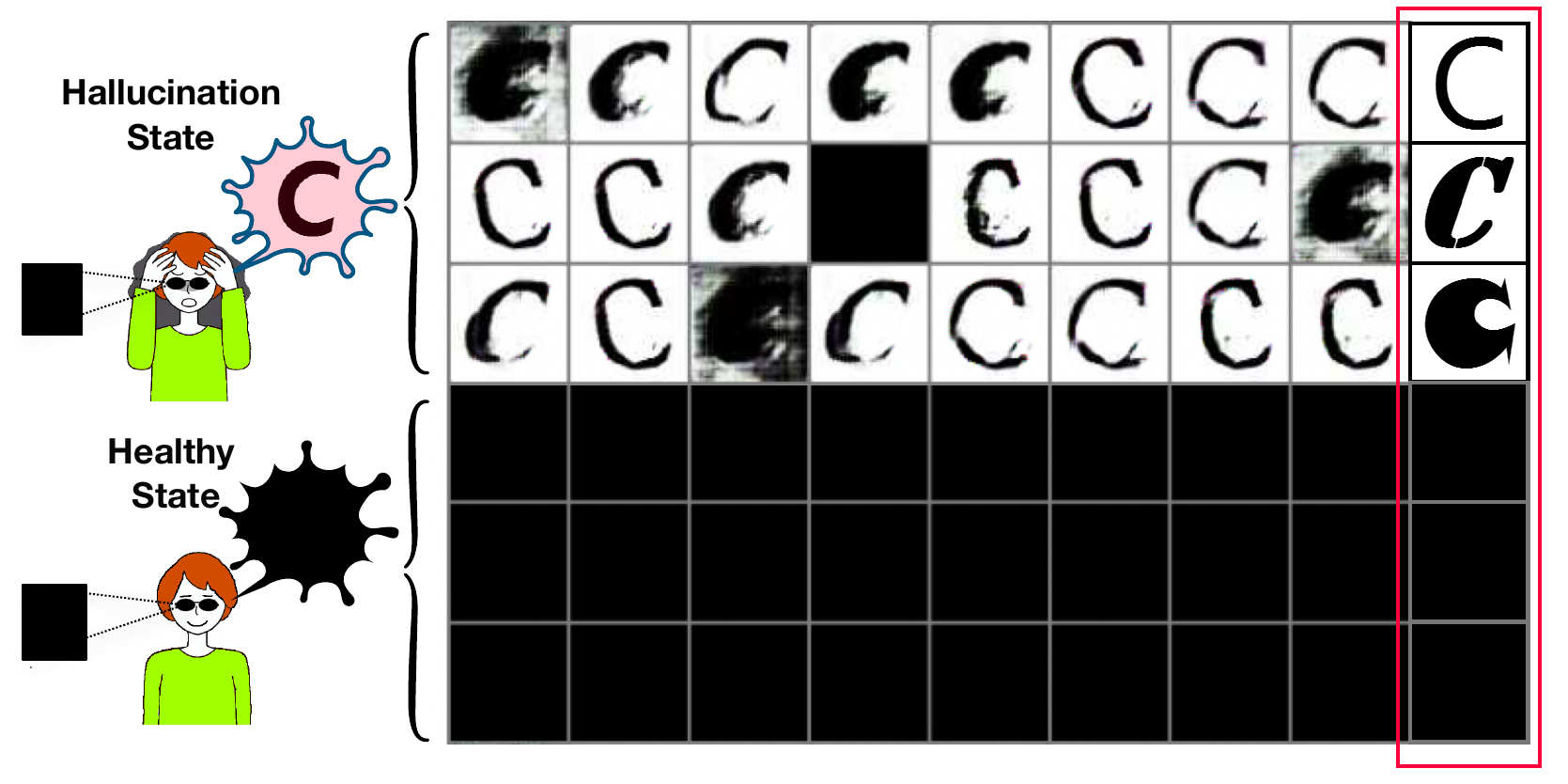}%
  \label{b_c}
}%
 \hfill%
	\subfloat[]{%
		\includegraphics[clip,width=0.38\columnwidth,height=0.33\linewidth]{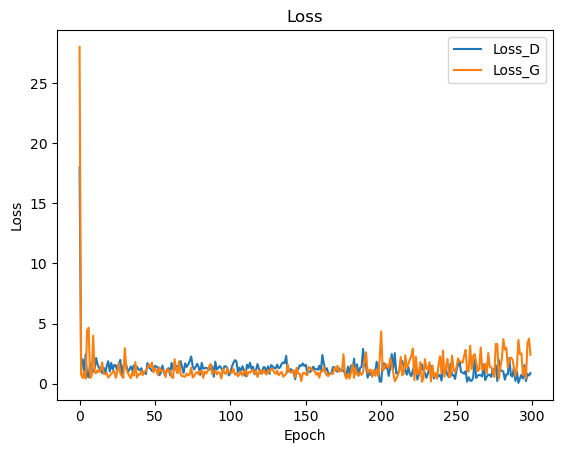}%
        \label{b_c_loss}
	}%
\end{center}
	\caption{
    Results of Hallu-GAN$^+$ with the Charles Bonnet syndrome data. The environmental input image of the generator is black. (a) Result of the Hallu-GAN$^+$  model. The first three rows of results depict a hallucinating visual system.  The last three rows illustrate a visual system at rest. It is important to note that columns 1 through 8 contain images generated by our model, while only the last column features randomly selected images from the training set. The character figures of the last column are under public dataset~\cite{kumar2018envisioned}. (b) Loss curves of  Hallu-GAN$^+$ during training with the Charles Bonnet syndrome dataset. The losses of the generator (G) and discriminator (D) stabilize after 300 epochs of training, reflecting the convergence of the proposed model.}
	\label{b_c_images}
\end{figure}

Fig.~\ref{b_c_loss} shows the expected loss during the training of the GAN model. Fig.~\ref{b_c} displays the images created using our suggested model for the hallucinating brain.
The last column in Fig.~\ref{b_c} displays a sample of randomly chosen photographs from the dataset, and the subsequent columns display the images produced by our model. The IS metric for the Hallu-GAN$^+$ is 2.64.

The results of the third experiment, i.e., Hallu-GAN$^+$ with EEG dataset of Charles Bonnet syndrome suggest that our model is able to represent the functioning of the visual system both during rest and during hallucinations.  
Our model attempts to generate the actual input image when it is at the rest state; but, when it is at the hallucination state, it attempts to generate a new image that is different from the input image. 
Due to the patient's blindness, the input element of the brain state indicates that other brain regions and prior memories may cause visual hallucinations. The patient's visual system uses prior memories and the output of other brain areas to reconstruct things that are not from the external environment. In this regard, we can utilize this model to analyze and detect the effective circumstances in which visual hallucinations arise. 
As a practical example, this model generates the hallucinogenic image in a hallucination state and the normal image in a resting state as a result of the data from Charles Bonnet's syndrome. 
In future work, we can enhance the model by categorizing more than two labels in order to analyze different contexts where hallucinations can occur. 

Our proposed models offer a basic comprehension of the functioning of the visual system in a brain experiencing hallucinations, as indicated by the results.
After the training step, the generator network of our model can generate the output with hallucination based on some conditions, a functionality similar to that in some areas of the brain.  Hence, the generator network formalizes the occipital lobe, visual cortex, and parietal area functionality in the hallucinating brain~\cite{ramirez2007cerebral, diederich2009hallucinations}. 
Additionally, the discriminator network in our model is capable of recognizing outputs that exhibit hallucination, resembling functions performed by certain brain regions. If the discriminator network is disrupted, it loses the ability to distinguish real data from hallucinated data. In this scenario, the Hallu-GAN model behaves in a manner akin to a hallucinating brain. Consequently, the discriminator encapsulates the functional roles of the prefrontal cortex and the inferior frontal gyrus.~\cite{gershman2019generative}. 

\section{Discussion}
\label{sec_Discussion}
Our current perspective focuses on the neurobiology of hallucinations from a modeling perspective.
Although the connection between GANs, mental disorders, and learning in the brain has been discussed previously~\cite{deperrois2022learning, deperrois2023learning, gershman2019generative}, our current model expands the discourse in various aspects.

While in~\cite{deperrois2022learning, deperrois2023learning}, a cortical architecture inspired by GANs with three states of learning is proposed to simulate the cortical learning process, we provided the Hallu-GAN and Hallu-GAN$^+$  models inspired by GAN for the hallucinatory visual system.  
Specifically, our study suggests and supports a mapping between the areas of a hallucinating brain and the elements within GANs. Neurologically, dopamine is crucial for reinforcing actions in behavioral learning, while other neuromodulators play roles in memory formation. Thus, neurotransmitters are essential for coordinated brain responses. Perturbations in neurotransmitter function, like in visual hallucinations, alter brain mechanisms and create adversarial interactions among brain areas.



In \cite{gershman2019generative}, an adversarial framework for probabilistic computation in the brain is proposed for analyzing some symptoms of mental disorders (such as delusions). A novel framework is proposed to explain intrusive experiences in PTSD, drawing inspiration from the generative adversarial process in machine learning~\cite{cushing2023generative}. This framework incorporates perceptual mechanisms and utilizes a cortical architecture similar to GANs to create a perceptual reality monitoring system.  In a similar context, in this paper, we introduced the interconnected areas of a hallucinating brain, which interact through an adversarial mechanism which can be effectively modeled by the Hallu-GAN model. Subsequently, we implemented and evaluated the Hallu-GAN model, utilizing Charles Bonnet's EEG data. Finally, we proposed and analyzed an extended version of this model, aiming to investigate both the healthy and hallucinatory visual systems. 

Our models suggest that when the visual cortex is damaged, it acts as a generator in a zero-sum game with reality-monitoring detector areas acting as a discriminator (similar to the GAN mechanism). 
Depending on the input image, the generator may produce a new image that appears to be real in an effort to deceive the discriminator. Hallucination happens when the discriminator is deceived. This enhances our understanding of the processes underlying hallucinations.

Before concluding this paper, it is worth mentioning that the proposed model can be used as a means to personalized medicine. In  particular, the model can be utilized to develop clinical decision support (CDS) applications for the clinician according the following guideline:
\begin{itemize}
    \item Preparing  EEG dataset per patient: this dataset includes EEG  of the hallucination and non-hallucination state of the person.
    \item  Preparing image dataset per patient: this dataset includes environment images and images based on the patient's hallucination (self-declared).
    \item Training the model with EEG and images in a hallucinatory and non-hallucinatory states.
    \item Testing and evaluating the model
\end{itemize}

After doing these steps, the neurologist/psychiatrist can use this application to detect hallucinogenic situations to prevent a patient from entering a hallucinatory state. 

\section{Conclusion}
\label{sec_Conclusion}

In the context of modeling functions of the human brain, we presented a model for the hallucinating brain. Focusing on visual hallucinations and some of its so far known neurological causes, we characterized an adversarial mechanism between different areas of the brain. We then showed how this adversarial setup can be modeled by GAN. In particular, we proposed exploiting the Hallu-GAN model. The proposed model can be viewed as the first steps of an addendum to the results of \cite{gershman2019generative}, providing evidence on how the idea of a generative adversarial brain can be extended to hallucinations as well.

Overall, we believe that our model provides several advantages in understanding hallucination-related disorders. It offers an explanation for why past experiences are subjectively experienced by individuals and why memories can be mistaken for current reality. By shedding light on these phenomena, our research contributes to a better understanding of hallucinations and opens up new directions for future studies in this field. Therefore, an interesting path for future work is to expand the generator to some subnetworks to better analyze the vision system. Finally, generalizing the proposed adversarial framework to other types of hallucinations would be intriguing.


Another potential avenue for future research would be to enhance the network architecture of our proposed model by drawing inspiration from biological networks and learning techniques. While CNNs are crucial for Hallu-GAN models, and can achieve human-like performance through experiential learning, they have limitations compared to visual systems.
 Weight sharing simplifies CNN training but does not align with biologically plausible visual feature selectivity. Additionally, the training process presents challenges for biological plausibility, as the backpropagation algorithm is not a suitable approximation of how the visual system learns. Gaining insight into how CNNs can be optimized to model rodent vision would greatly enhance our understanding of the distinctions between primate and rodent vision. This would also enable the utilization of exploration strategies outlined in this context for studying rodent vision. 
Indeed, these challenges add to the beauty of this field of research and present an opportunity for further exploration.

%

\section*{Acknowledgements}
The authors would like to thank Dr. Charlotte Martial, for providing the clinical data, and also for her fruitful  comments.

 \section*{Competing interests}
 The authors declare no competing interests.
\bibliography{sample}








\end{document}